\documentclass[aps,twocolumn]{revtex4-1}

\begin{document}

%\preprint{}

\title{Ho{\u r}ava-Lifshitz Gravity on 
Primordial Black Holes }

\author{Hikoya Kasari}
\email[]{kasari@keyaki.cc.u-tokai.ac.jp}
\author{Takehiko T. Fujishiro}
\email[]{fuji46@keyaki.cc.u-tokai.ac.jp}

%\thanks{}
\affiliation{Department of Physics, School of Science, Tokai University, 
Hiratsuka Kanagawa 259-1292, Japan}

\date{\today}

\begin{abstract}
We tested the Ho{\u r}ava Lifshitz (HL) quantum gravity model 
by using the 
L{\" u}, Mei and Pope solutions for primordial black holes (PBHs) and 
the observational upper bounds of the PBH density parameters. 
We found that, although the HL model is severely constrained, 
it is not ruled out. 
When our analysis is combined with that of Dutta and Saridakis 
the observed value of the density parameter $\Omega_{PBH}$ 
might rise by several percent as the running energy parameter 
$\lambda$ increases.
\end{abstract}

%\keywords{}

\maketitle

\section{Introduction\label{Introduction}}

Ho{\u r}ava\cite{Horava 3} has applied the anisotropic scaling that Lifshitz developed\cite{lifshitz} for quantum gravity and established the power-counting renormalizability of the theory 
in 3+1 dimensional space-time. 
This model is usually called Ho{\u r}ava Lifshitz (HL) gravity. 
Cosmological 
solutions of HL gravity
for the Friedmann-Lema{\^ i}tre-Robertson-Walker metric have been described L{\" u}, Mei and Pope (LMP) 
\cite{LMP}. 
Although there has also been a lot of other work on HL gravity \cite{otherBHsolution}, the LMP solution is the most basic one, and therefore our discussion is based around it.
The LMP solutions are based on conditions of 
relaxed projectability 
and of detailed balance. 
Recently, Gong et al.\cite{sasaki} have shown 
using linear cosmological perturbation theory 
with projectability that HL gravity is 
reduced by the Faddeev and Jackiw condition\cite{Faddeev} and 
that there is a scalar ghost mode for $\frac{1}{3} < \lambda < 1$,
where $\lambda$ is the running energy parameter in the HL model. 

In 1971, Hawking showed that black holes should have been 
generated with various masses in the early universe\cite{pbh}.
Nowadays, such black holes are called primordial black holes 
(PBHs), and their masses $M$ are estimated from the Hubble equation as 
$M 
\simeq 10^{15} \left( \frac{t_f}{10^{-23}} \right) g$,
where 
$t_f$ is the current time from the big bang.
The black holes emit energy according to the Stefan-Boltzmann law 
and evaporate.
Because the time necessary for evaporation is 
$t_{ev} 
\simeq 10^{10} \left( \frac{M}{10^{15}} g \right)^3 $ years, 
black holes that are lighter than $10^{15} g$ have completely evaporated via Hawking radiation
before the present $t_f(\simeq 10^{10} yr)$\cite{PDG} and only heavier PBHs remain. 
In \cite{obHR}, the observational values are given 
between 35 MeV and 175 MeV for $\gamma$-rays, 
$\Omega_{PBH} \leq (7.6 \pm 2.6) \times 10^{-9} h^{-2} 
 \simeq 1.4 \times 10^{-8}$,
with the Hubble expansion rate $h=0.73^{+0.04}_{-0.03} \approx 0.73$, 
and between 30 MeV and 120 GeV for $\gamma$-rays,
$\Omega_{PBH} \leq (5.1 \pm 1.3) \times 10^{-9} h^{-2}  
 \simeq 9.5 \times 10^{-9}$.
The observational upper limit on the PBH density parameter from measurements of Hawking radiation is \cite{obHR}
\begin{equation}
\Omega_{PBH} \leq 10^{-8}.
\label{eqn:upperbound}
\end{equation}
However, PBHs are as yet unidentified stellar objects\cite{obPHB2005} and may constitute all the dark matter\cite{yanagida}. Regrettably, the nature of dark matter and dark energy 
is not yet understood, but 
here we assume that the density parameter of the PBH $\Omega_{PBH}$ contains the density parameters of the dark energy $\Omega_{DE}$ or the dark matter $\Omega_{DM}$.
The values of the various density parameters from WMAP \cite{Omega7yr} are $\Omega_{b} \simeq 0.0449 \pm 0.0028$ for baryons, 
$\Omega_{DM} \simeq 0.222 \pm 0.026$ for the dark matter
 and 
$\Omega_{DE} \simeq 0.734 \pm 0.029$ for the dark energy.
 
In HL gravity the (running) energy parameter $\lambda$ lies in 
the range $\frac{1}{3} \leq \lambda \leq 1$. 
The present value of the energy parameter is $\mid \lambda_0 - 1 \mid \leq 0.002$, 
based on observations of supernova, 
the baryon acoustic oscillation (BAO) 
and the cosmic microwave background (CMB) \cite{DS10}.
However, because we are only discussing PBHs 
and the focus 
of our research 
is different 
from theirs, such a tight bound on $\lambda_0$ may not be appropriate. 
Note also that many works have been carried out that focus on dark energy 
in the HL model\cite{DE}. 

This rest of this article is arranged as follows. 
In the next section we review HL gravity and the LMP solutions. 
In the following section, we 
compare the LMP solutions with current observational PBH data.
Finally, we present our conclusions. 
The values in SI units of some cosmological constants are given in an appendix.

\section{Ho{\u r}ava Lifshitz and LMP solutions}
\label{HL}

HL gravity\cite{Horava 3} has the anisotropic scaling property $x \rightarrow b x, \ \ t \rightarrow b^{z} t $, with critical exponent $z$. The theory is Lorentz invariant 
in the long distance, IR limit ($z=1$)
but it is not Lorentz invariant 
in the short distance, UV limit ($z=3$).

The Arnowitt-Deser-Misner (ADM) 
decomposition of 4-dimensional space-time is given by
\begin{eqnarray}
ds^2 &=& c^2 N^2 dt^2 + g_{ij} 
\left( dx^i - c N^i dt \right)
\left( dx^j - c N^j dt \right),
\label{eqn:ADM}
\end{eqnarray}
where $N$ is the lapse field, $N^i$ is the shift field 
and $g_{ij}$ is the spatial metric.
The scale dimensions of physical quantities in units of mass are
$\left[ G_N \right] = -2$, 
$\left[ N \right] = [g_{ij}] = 0$, and 
$\left[ N_i \right] = z - 1$.
For simplicity, we assume that $N$ depends only on time $t$ 
($i.e.$, projectability). 
The ADM decomposition of Einstein-Hilbert (EH) action is 
\begin{equation}
S_{EH}= \frac{1}{16 \pi G_N} \int d^4 x \sqrt{g} N 
\left\{ \left( K_{ij} K^{ij} - K^2 \right) + R - 2 \Lambda \right\},
\label{eqn:EH}
\end{equation}
where $G_N$ is Newton's gravitational constant and $K_{ij}$ 
is the extrinsic curvature defined by 
$K_{ij} = \frac{1}{2N} 
\left( {\dot g}_{ij} -  \nabla_i N_j -  \nabla_j N_i \right)$.
The original action of Ho{\u r}ava\cite{Horava 3} is 
\begin{eqnarray}
S_{HL} &=& \int dt d^3{\bf x} ({\mathcal L}_0 + {\mathcal L}_1), 
\label{eqn:HLaction}
\\ 
{\mathcal L}_0 &=& {\sqrt g} N \left\{ \frac{2}{\kappa^2} 
\left( K_{ij}K^{ij} - \lambda K^2 \right) \right. \nonumber \\ 
 & & \left.
+ \frac{\kappa^2 \mu^2 
        \left( \Lambda_W R - 3 {\Lambda_W}^2 \right)}
       {8 (1-3\lambda)}
\right\}, 
\label{eqn:L0}
\\
{\mathcal L}_1 &=& {\sqrt g} N \left\{ 
\frac{\kappa^2 \mu^2 (1-4\lambda)}{32(1-3\lambda)} R^2 
- \frac{\kappa^2}{2 w^4} 
\left( C_{ij} - \frac{\mu w^4}{2} R_{ij} \right)
 \right. \nonumber \\ 
& & 
\left. 
\times 
\left( C^{ij} - \frac{\mu w^4}{2} R^{ij} \right) 
\right\},
\end{eqnarray}
where $\lambda, \kappa, \mu, w$ and $\Lambda_W$ are constant parameters, and
$C_{ij}$ is the Cotton tensor 
$C^{ij} = \varepsilon^{ikl} \nabla_k 
\left( R^j_l - \frac{1}{4} R \delta^j_l \right)$.
Note that only $w$ controls the strength of the interaction and 
the free state corresponds to the limit $w \rightarrow 0$. Also note that 
we can identify $\lambda \rightarrow 1$ as the IR limit 
and $\lambda \rightarrow 1/3$ as the UV limit.
By comparing 
 (\ref{eqn:HLaction}) with (\ref{eqn:EH}) the following relations are obtained. 
\begin{equation}
c=\frac{\kappa^2 \mu}{4} \sqrt{\frac{\Lambda_W}{1-3\lambda}},
\ \ \ 
G_N=\frac{\kappa^2}{32 \pi c},
\ \ \ 
\Lambda = \frac{3}{2} \Lambda_W,
\label{eqn:CGL}
\end{equation}
where $\Lambda$ is the cosmological constant.
In 3+1 dimensions, the correspondence of the HL action to 
the Einstein-Hilbert (EH) action demands $\lambda =1$ ($i.e.$, the IR limit).

If we demand that the speed of light 
is positive in (\ref{eqn:CGL}), then $\Lambda_W$ is negative.
The analytic continuation 
$\mu \rightarrow i\mu$ and $w^2 \rightarrow -iw^2$ yields 
positive $\Lambda_W$. 
Then, we can describe the speed of light as
$ c= \frac{1}{4} \kappa^2 \mu \sqrt{\frac{\Lambda_W}{3\lambda-1}}$
  \ for $\lambda > \frac{1}{3}$.

The Friedmann-Lema{\^ i}tre-Robertson-Walker form is
\begin{equation}
ds^2 = - c^2 dt^2 + a^2(t) \frac{dr^2}{1 - K r^2}+ r^2 \left(
d \theta^2 + \sin^2 \theta d \phi^2 \right).
\end{equation}
where $K= -1, 0, 1$ corresponds to a closed, flat or open 
universe, respectively, 
and $a(t)$ is the scale factor.
For this metric the Friedmann equation (in units where $c=1$) is
\begin{eqnarray}
\left( \frac{{\dot a}}{a} \right)^2  &=& 
\frac{2}{3\lambda-1} \nonumber \\ 
 & & \times 
\left( \frac{\Lambda_W}{2} + \frac{(8 \pi G_N) \rho}{3} 
- \frac{K}{a^2} + \frac{K^2}{2 \Lambda_W a^4} \right), 
\label{eqn:LMPfriedmann}
\end{eqnarray}
where $\rho = \rho_m + \rho_r$ is the sum of the energy density of matter ($\rho_m$)
and the energy density of radiation ($\rho_r$).

\section{Comparison with PBH data}
\label{PBHdata}

Reintroducing the speed of light explicitly, 
the density parameter is defined by the Friedmann equation 
(\ref{eqn:LMPfriedmann}) through
\begin{eqnarray}
1 &=& \frac{2 (8\pi G_N)c}{3(3\lambda-1)H^2} \rho_m + 
\frac{2(8\pi G_N) c}{3(3\lambda-1)H^2} \rho_r + 
\frac{2 c^2 K}{(3\lambda-1) H^2 a^2} \nonumber \\ 
 & &-\frac{c^2}{(3\lambda-1) H^2} \left[ \frac{K^2}{\Lambda_W a^4} + 
 \Lambda_W  \right],
\label{eqn:Friedmanneq}
\end{eqnarray}
where the energy density of matter $\rho_m$ consists of   
the baryonic energy density $\rho_b$ and 
the dark matter energy density $\rho_{DM}$,  
the total energy density of radiation $\rho_r$ consists of 
the photon energy density $\rho_{\gamma}$ and 
the muonic energy density $\rho_{\mu}$ 
($i.e.$,
$\rho_r= \rho_{\gamma} + \rho_{\mu} = 1.69 \rho_{\gamma}$\cite{KolbTurner}
), 
and $H=\frac{\dot a}{a}$ is the Hubble parameter.
On the right hand side of (\ref{eqn:Friedmanneq}), 
we identify 
the first term as the matter density parameter $\Omega_{m}$, 
the second term as the radiation density parameter $\Omega_{r}$ 
and the third term as the curvature density parameter $\Omega_{K}$.
Consequently we obtain
\begin{equation}
1-\Omega_m - \Omega_r-\Omega_K 
= -\frac{c^2}{(3\lambda-1) H^2} \left[ \frac{K^2}{\Lambda_W a^4} + 
 \Lambda_W  \right].
\end{equation}
The right hand side of this equation is called the dark radiation
\cite{HLcosmology1,HLcosmology2}.
The dark radiation contributes negative energy density and the greater part of this 
contribution disappears if the curvature $K$ is zero.

We make the following assumptions. 
The density parameter of PBHs takes its maximum value
$\Omega_{PBH,max}$ if it includes 
the density parameters of the dark matter 
$\Omega_{DM}$ and 
dark energy 
$\Omega_{DE}$. So 
\begin{equation}
\Omega_{PBH,max} \leq 0.96 = 
\Omega_{DE} + \Omega_{K} + \Omega_{DM} + \Omega_{r}. 
\label{eqn:OmegaPBHmax}
\end{equation}
From current observations (via Hawking radiation)\cite{10^-8} 
(also see \cite{obHR,obPHB2005}), 
it is thought that the present density parameter for PBHs 
$\Omega_{PBH_0}$ satisfies 
$\Omega_{PBH_0} \leq 10^{-8}$ (\ref{eqn:upperbound}).
Because (\ref{eqn:upperbound}) is based on data from radiation observations, 
$\Omega_{PBH_0}$ may contain the dark radiation term 
($i.e., \Omega_{DE}$).
Thus, we obtain 
\begin{eqnarray}
\Omega_{PBH} &\simeq& 
-\frac{c^2}{(3\lambda-1)H^2} 
\left( \frac{K^2}{\Lambda_W a^4} + \Lambda_W \right), 
\label{eqn:DPPBH} \\ 
\Omega_{PBH_0} &\simeq& 
-\frac{c_0^2}{(3\lambda_0-1)H_0^2} 
\left( \frac{K_0^2}{\Lambda_{W_0} a_0^4 } + \Lambda_{W_0} \right). 
\label{eqn:DPPBH0} 
\end{eqnarray}
Because $\Lambda_W$ is small, we can neglect second terms of 
 (\ref{eqn:DPPBH}) and (\ref{eqn:DPPBH0}).
Then, assuming $K_0 \simeq K, \ 
\Lambda_{W_0} \simeq \Lambda_{W}$ and $  a_0 = 1$, 
\begin{eqnarray}
\Omega_{PBH} &=&  -\frac{c^2}{(3\lambda-1)H^2} 
 \frac{K^2}{a^4}  \sim 0.96, 
\label{eqn:sDPPBH} \\ 
\Omega_{PBH_0} &=&  -\frac{c_0^2}{(3\lambda_0 -1)H_0^2} 
 K^2 \sim 10^{-8}. 
\label{eqn:sDPPBH0} 
\end{eqnarray}
When we take the ratio of (\ref{eqn:sDPPBH}) and (\ref{eqn:sDPPBH0}),
we obtain 
\begin{equation}
\frac{\Omega_{PBH}}{\Omega_{PBH_0}} = 
\frac{3 \lambda_0 -1}{3 \lambda -1}
\frac{H_0^2 c^2}{H^2 a^4 c_0^2}
\sim 0.96 \times 10^8 \sim 10^8.
\label{eqn:ratioofdensityparameter}
\end{equation}
Substituting 
$\left( \frac{c}{c_0} \right)^2 = \frac{3 \lambda_0 - 1}{3 \lambda - 1}$
into (\ref{eqn:ratioofdensityparameter}) gives 
\begin{equation}
\frac{H_0^2}{H^2 a^4} 
\left( \frac{3 \lambda_0 - 1}{3 \lambda - 1} \right)^2
\simeq 10^{8}.
\end{equation}
Consequently,
\begin{eqnarray}
H_0^2 &\sim& 10^8 H^2 a^4  
\left( \frac{3 \lambda - 1}{3 \lambda_0 - 1} \right)^2  \nonumber \\ 
 &=& 10^8 {\dot a}^2 a^2 
 \left( \frac{3 \lambda - 1}{3 \lambda_0 - 1} \right)^2.  
\label{eqn:108}
\end{eqnarray}
Introducing $ r^2 \equiv 10^8 $, we have 
\begin{eqnarray}
H_0^2 &=& r^2 {\dot a}^2 a^2 
\left( \frac{3 \lambda - 1}{3 \lambda_0 - 1} \right)^2.  
\end{eqnarray}

From (\ref{eqn:LMPfriedmann}) 
the scale factor is 
\begin{eqnarray}
\left( \frac{{\dot a}}{a} \right)^2  
 &=& \left( \frac{\Lambda_W}{3\lambda-1} 
           + \frac{2(8 \pi G_N) \rho}{3 (3\lambda-1)} \right) \nonumber \\ 
 & & 
    - \frac{2 K}{(3\lambda-1)} \frac{1}{a^2} 
    + \frac{K^2}{(3\lambda-1) \Lambda_W} \frac{1}{a^4}, 
\label{eqn:scale_factor}
\end{eqnarray}
where $\rho= \rho_m + \rho_r.$
Now, we define 
\begin{eqnarray}
A  &\equiv&  \frac{\Lambda_W}{3\lambda-1} 
           + \frac{2(8 \pi G_N) \rho}{3 (3\lambda-1)},  \\ 
B &\equiv& \frac{2 K}{(3\lambda-1)}, \\ 
C &\equiv& \frac{K^2}{(3\lambda-1) \Lambda_W} 
\end{eqnarray}
with the present values being denoted by the subscript $0$. 
Then (\ref{eqn:scale_factor}) becomes
\begin{eqnarray}
{\dot a}^2 = A a^2 - B + C \frac{1}{a^2}, 
\label{eqn:adiffeq}
\end{eqnarray}
which can be solved easily. 
This differential equation has two types of solution 
for the present Hubble parameter $H_0$:
\begin{eqnarray}
H_{0}^{(I)} &=& - r \frac{e^{-2 \sqrt{A_0} t_0}}{4 \alpha_0^2 \sqrt{A_0}}
\left(  B_0^2 - 4 A_0 C_0 - \alpha_0^4 e^{4 \sqrt{A_0}t_0}\right)
 \nonumber \\ 
 & & \times 
\left( \frac{3 \lambda - 1}{3 \lambda_0 -1} \right), \\ 
H_{0}^{(II)} &=& r \frac{e^{2 \sqrt{A_0} t_0}}{4 \beta_0^2 \sqrt{A_0}}
\left( B_0^2 - 4 A_0 C_0 - \beta_0^4 e^{- 4 \sqrt{A_0}t_0}\right)
 \nonumber \\ 
 & & \times 
\left( \frac{3 \lambda - 1}{3 \lambda_0 -1} \right), 
\end{eqnarray}
where $\alpha_0$ and $\beta_0$ are the constants of integration. 
With assumption that the extrinsic curvature $K$ is zero, we have 
$B_0 = C_0 = 0$. Then the constants of integration are given by 
\begin{eqnarray}
\alpha_0|_{K=0} &=& \pm 2 e^{- \sqrt{A_0}t_0} \sqrt{A_0}, \\ 
\beta_0|_{K=0} &=& \pm 2 e^{ \sqrt{A_0}t_0} \sqrt{A_0} 
\end{eqnarray}
and the present Hubble parameters are 
\begin{eqnarray}
H_{0}^{(I)}|_{K=0} &=& r \sqrt{A_0} 
\left( \frac{3 \lambda - 1}{3 \lambda_0 -1} \right), \\ 
H_{0}^{(II)}|_{K=0} &=& - r \sqrt{A_0} 
\left( \frac{3 \lambda - 1}{3 \lambda_0 -1} \right).
\end{eqnarray}
That is, 
\begin{eqnarray}
H_0^2 &=& 
r^2 A_0 \left( \frac{3 \lambda - 1}{3 \lambda_0 -1} \right)^2 
\nonumber \\ 
 &=& r^2 
\left\{ \frac{\Lambda_W}{3 \lambda_0 -1} 
+ \frac{2 (8 \pi G_N)}{3 (3 \lambda_0 -1)} \rho_0 \right\} 
\nonumber \\ 
 & & \times 
\left( \frac{3 \lambda - 1}{3 \lambda_0 -1} \right)^2,
\end{eqnarray}
where 
\begin{eqnarray}
\rho_0 &=& \rho_{m_0} + \rho_{\gamma_0} \simeq \rho_{m_0} 
 = 0.24 \times \frac{3 H_0^2}{8 \pi G_N}.
\end{eqnarray}
Then $H_0^2$ becomes 
\begin{equation}
H_0^2 = \frac{r^2 \left( 3 \lambda -1 \right)^{2}}{
\left( 3 \lambda_0 -1 \right)^3  }
\left\{ \Lambda_W + 0.48 H_0^2 \right\}.
\end{equation}
With $\Lambda_W = 7.77 \times 10^{-36} [1/s^2]$ 
and $H_0 = \frac{1}{4.35 \times 10^{17} [s]}$ (see Appendix), we obtain 
\begin{eqnarray}
\frac{(3\lambda_0 -1)^3}{r^2 (3 \lambda -1)^2} &=& 
 \frac{\Lambda_W}{H_0^2} + 0.48 
 = 1.95.
\end{eqnarray}
Thus,
\begin{equation}
\lambda = \frac{1}{3} + \frac{1}{3} \sqrt{ 
\frac{(3 \lambda_0 -1)^3}{1.95 \times r^2} }.
\end{equation}
In the HL theory the energy parameter $\lambda$ is in the range 
$\frac{1}{3} \leq \lambda \leq 1$, so we obtain 
\begin{equation}
\frac{1}{3} \leq \lambda_0 \leq \frac{1+ \sqrt[3]{7.80 \times r^2}}{3}.
\label{eqn:ourresult}
\end{equation}
The physical interpretation of 
(\ref{eqn:ourresult}) 
is discussed in the next section.

\section{Conclusions}
\label{Conclusion}

If 0.96 is substituted for $\Omega_{PBH,max}$ and 
$10^{-8}$ is substituted for $\Omega_{PBH_0}$, 
(\ref{eqn:ourresult}) becomes 
\begin{equation}
\frac{1}{3} \leq \lambda_0 \leq 307. 
\label{eqn:10to-8rengeoflambda}
\end{equation}
Although this range contains 
the parameter range ($\frac{1}{3} \leq \lambda_0 \leq 1$) of the HL model, it is clear that HL gravity is considerably more restrictive.

Inversely, when the limitations of the HL model are imposed on 
(\ref{eqn:ourresult}) 
we obtain
\begin{equation}
\Omega_{PBH_0} \simeq \Omega_{PBH,max}\;.
\label{eqn:relationoandmax}
\end{equation}
This is consistent with the result of Frampton et al.\cite{yanagida}.

If the black holes do not contribute to the density parameter(s), 
the present energy parameter $\lambda_0$ 
in (\ref{eqn:ourresult}) 
becomes 1/3 (i.e., we attain the UV limit). 
In the HL theory the general theory of relativity (GR) is not formed 
in the UV limit, but is recovered in the IR limit. 
Thus, $\lambda_0 \simeq 1/3$ is inconsistent with the status of GR of today.

It is interesting to combine our analysis with that of Dutta and Saridakis\cite{DS10}. 
If (\ref{eqn:ourresult}) is combined with their result 
$|\lambda_0 -1|< 0.002$, it becomes 
\begin{equation}
1.02 \times \Omega_{PBH_0} < \Omega_{PBH,max} < 1.03 \times \Omega_{PBH_0}.
\end{equation}
This shows that $\Omega_{PBH,max}$ can be larger than present observed value 
$\Omega_{PBH_0}$ by few percent. 
This is consistent with the prediction in the HL theory that 
the horizon of a black hole appears as $\lambda$ increases 
\cite{Horava 3} \cite{LMP}.
A PBH that has not been observed up to now 
will be observed in the future because of an increase in $\lambda$.

We thus arrive at the following three possible conclusions. 
(1) If the HL theory and the current observational upper bound 
$\Omega_{PBH_0} \leq 10^{-8}$ are correct, 
then the assumption $\Omega_{PBH,max} \simeq \Omega_{DE/DM}$
is doubtful. 
(2) If the current observational upper bound $\Omega_{PBH_0} \leq 10^{-8}$ and 
the assumption $\Omega_{PBH,max} \simeq \Omega_{DE/DM}$ 
are correct, 
then the HL model is doubtful.
(3) If HL gravity and 
the relation (\ref{eqn:relationoandmax}) are correct, 
then 
the observational $\Omega_{PBH_0}$ may 
increase slightly 
as $\lambda$ increases.

\appendix*
\section{Cosmological constants} 
\label{appendix}
The present Hubble parameter is observed to be
$H_0 = 71.0 \pm 2.5 [km/(s M_{pc})]$ \cite{Omega7yr}. So we take
$H_0 = 71.0 \left[ \frac{km}{s M_{pc}} \right] 
 = 2.30 \times 10^{-18} \left[ \frac{1}{s} \right]  
 = \frac{1}{4.35 \times 10^{17} [s]}$,
where $M_{pc} = 3.09 \times 10^{22} [m] $.

Generally the cosmological constant, 
$i.e.$, the density parameter of 
the vacuum (dark) energy, is given by 
$\Omega_{DE} = \frac{\Lambda}{3 H_0^2} = 0.734$. %\nonumber
So we obtain 
$\Lambda = 3 \times 0.734 \times H_0^2  = 
1.16 \times 10^{-35} \left[ \frac{1}{s^2} \right]$ 
and 
$\Lambda_W = \frac{2}{3} \Lambda 
 = 7.77 \times 10^{-36} \left[ \frac{1}{s^2} \right]$.


\end{document}